\documentclass[preprint]{elsarticle}

\usepackage{lineno,hyperref}
\modulolinenumbers[5]

\journal{Applied Soft Computing}

\usepackage[utf8]{inputenc}
\usepackage{amsmath}
\usepackage{algorithm}
\usepackage{algorithmic}
\usepackage{stfloats}
\usepackage[table]{xcolor}

\begin{document}

\begin{frontmatter}

\title{Development of a hybrid method for stock trading based on TOPSIS, EMD and ELM}

\author[ref1,ref2]{Elivelto Ebermam}
\author[ref2]{Helder Knidel}
\author[ref1,ref2,ref3]{Renato A. Krohling}

\cortext[mycorrespondingauthor]{Email addresses: eliveltoebermam@gmail.com (Elivelto Ebermam), helkni@gmail.com (Helder Knidel), krohling.renato@gmail.com (Renato A. Krohling)}

\address[ref1]{Graduate Program in Computer Science, PPGI, UFES - Federal University of Espírito Santo, Av. Fernando Ferrari 514, Vitória CEP: 29060-270, Brazil}
\address[ref2]{Laboratory of computing and engineering inspired by nature, UFES - Federal University of Espírito Santo, Labcin, Av. Fernando Ferrari 514, Vitória CEP: 29060-270, Brazil}
\address[ref3]{Production Engineering Department, UFES - Federal University of Espírito Santo, Av. Fernando Ferrari 514, Vitória CEP: 29060-270, Brazil}

\begin{abstract}
Deciding when to buy or sell a stock is not an easy task because the market is hard to predict, being influenced by political and economic factors. Thus, methodologies based on computational intelligence have been applied to this challenging problem. In this work, every day the stocks are ranked by technique for order preference by similarity to ideal solution (TOPSIS) using technical analysis criteria, and the most suitable stock is selected for purchase. Even so, it may occur that the market is not favorable to purchase on certain days, or even, the TOPSIS make an incorrect selection. To improve the selection, another method should be used. So, a hybrid model composed of empirical mode decomposition (EMD) and extreme learning machine (ELM) is proposed. The EMD decomposes the series into several sub-series, and thus the main component (trend) is extracted. This component is processed by the ELM, which performs the prediction of the next element of component. If the value predicted by the ELM is greater than the last value, then the purchase of the stock is confirmed.
The method was applied in a universe of 50 stocks in the Brazilian market. The selection made by TOPSIS showed promising results when compared to the random selection and the return generated by the Bovespa index. Confirmation with the EMD-ELM hybrid model was able to increase the percentage of profit tradings.
\end{abstract}

\end{frontmatter}


\section{Introduction}

In stock markets are negotiated  a large amount of money in many countries every business day. However, financial markets are essentially dynamic, non-linear, chaotic in nature \cite{tan2007}. Predicting the price or movement of stocks, as well as the best time to buy and sell them, is a difficult and open issue. This is due to the fact that the market behavior is volatile and influenced by several macroeconomic factors such as: political events, company policies, general condition of the economy, investors expectations, institutional investors choices, influence of other markets, investors psychology, among others \cite{vargas2017,tan2007}.

The predictability or non-predictability of the financial market has been the subject of discussion between researchers and professional investors for decades. One of the central points of this discussion is the validity of the efficient market hypothesis (HME), created by Fama in the 1960s. This hypothesis claims that price always incorporates all available information \cite{fama1970}. Thus, if information is reflected immediately in stock prices, then tomorrows price change is independent of current price changes, depending only on tomorrow's news \cite{malkiel2003}. News, however, are by definition unpredictable, making price changes also unpredictable and random \cite{malkiel2003}. This means that by this hypothesis market prices follow a random walk, and it is impossible to use past behavior to predict price of the assets \cite{papacostantis2009}.

However, many economists and statisticians believe that stock prices are at least partially predictable \cite{malkiel2003}. So, several works appeared aiming to predict the stock market and question the HME. Computational intelligence methods such as neural networks, evolutionary algorithms, fuzzy logic, among others, have been applied to this problem. Artificial neural networks have been relatively successful in predicting prices and movements in the stock market. Martinez et al. \cite{martinez2009} used a standard neural network to forecast the daily minimum and maximum prices for two stocks of the Brazilian stock market. Oliveira et al. \cite{oliveira2011} predict the closing price of the Petrobras stock (PETR4). Kara, Boyacioglu and Baykan \cite{kara2011} used neural networks and support vector machine to predict the movement of the Istanbul Stock Exchange Index (ISE). Nelson, Pereira and Oliveira \cite{nelson2017} used a long short-term memory (LSTM) to predict the price trend of five stocks in the Brazilian market. Vargas, Lima and Evsukoff \cite{vargas2017} combined convolutional neural network and LSTM to predict the movement of the S\&P500 index taking as input news headlines and technical indicators. Nascimento et al. \cite{nascimento2017} used VGRAM weightless neural networks to evaluate investment opportunities in the Brazilian futures market in the scale of tens of seconds.

Fuzzy logic has also been applied to financial market problems, mainly because of its ability to deal with uncertainties. Decision support systems, combining technical indicators and fuzzy logic, are used in \cite{dourra2002}, \cite{cheung2007} and \cite{ijegwa2014}. Recently, Naranjo, Arrayo and Santos \cite{naranjo2018} developed a stock trading system based on graphical patterns of candles using fuzzy logic.

Evolutionary algorithms have been used mainly to optimize parameters of technical indicators \cite{fernandez2008}, \cite{bodas2009}, \cite{fayek2013} and \cite{Almeida2016}. Brasileiro et al. \cite{brasileiro2013} developed a trading system using k-neighbors (KNN) algorithm, an indicator of technical analysis and optimization by colony of artificial bees.

Most of the works select an a priori stock or index to apply a certain method. However, selecting a stock to be traded is already a problem in itself, since each market has several stocks and choosing among them is not a trivial task. For this purpose, multi-criteria decision-making methods can be used. They have been successfully applied to the selection of alternatives in several types of applications, such as: property price evaluation \cite{krohling2011}, suppliers selection \cite{jadidi2010, deswal2015}, energy planning \cite{loken2007}, among others. A widely used technique is TOPSIS (Technique for Order Preference by Similarity to Ideal Solution), which evaluates the performance of the alternatives based on the similarity to the ideal solution \cite{yoon1981}.
  
Multi-criteria decision-making methods need to specify criteria based on market indices to buy or sell stocks. For the selection of stocks in the market, most of the works use fundamentalist criteria, that is, metrics based on the financial statements of companies. King, Jung and Cho \cite{kim2009} applied TOPSIS to asset selection in the Korean stock market. Bulgurcu \cite{bulgurcu2012} used TOPSIS to evaluate technology companies in the Istanbul stock market. Janani, Ehsanifar and Bakhtiarnezhad \cite{janani2012} used a variation of TOPSIS to form an investment portfolio and stock ranking for the Tehran stock market. Saleh and Kimiagari \cite{Kimiagari2017} used fuzzy TOPSIS to rank stocks in the same market. For the Brazilian market, Costa and Junior \cite{costa2013} pre-selected stocks with TOPSIS using as criteria: financial indices, a risk measure and a qualitative parameter of corporate governance. All the works cited previously were based on fundamentalist criteria.

However, fundamentalist analysis is more suitable for long-term investors, since the financial statements of companies are only disclosed from time to time (interval of a few months). Short-term investors typically use technical analysis. The technical analysis consists of analyzing the behavior of a stock in a financial market through the use of graphs and indicators in order to predict the future trend of prices \cite{murphy1999}. Technical analysis is commonly used through technical indicators, which are calculated based on the price history. Its advantage is that prices are available at any time and hence the indicators as well.

So, for short-term investments, the combination of a multi-criteria decision-making method such as TOPSIS and criteria derived from technical analysis may be more appropriate. However, TOPSIS only ranks the most suitable stocks to invest and there are days when most of the stocks have a price decrease, and therefore it is very likely that the choice of TOPSIS is incorrect on that day. In this way, using another technique together may improve the reliability of the investment, that is, the realization of the purchase or not of a certain stock.

Methodologies combining time series decomposition techniques and artificial neural networks have been successful in applications such as: wind speed prediction \cite{guo2012}, flow of passenger \cite{wei2012},  the use of solar energy \cite{majumder2017}, currency conversion \cite{das2017} among others. The specific combination of empirical mode decomposition (EMD) and  extreme learning machine (ELM) is used in \cite{lian2013}, \cite{majumder2017} and \cite{das2017}. 

The empirical mode decomposition technique can be used to perform the decomposition of the stock price series selected by TOPSIS and extract the trend component. Next, the extreme learning machine can be used to predict the trend, that is, whether it will rise or fall. By doing so, there is a confirmation (or not) that the selected stock should be purchased.

The main goal of this work is to construct a method capable of analyzing several stocks at the same time and deciding which stock should be purchased at each day. In addition, the method must be applicable in the real world. So, our aim is to carry out the selection of stocks for short-term purchase. In this way, the use of technical analysis indicators as criteria and the TOPSIS technique for decision making will be investigated. Next, to confirm the purchase of the stock selected by TOPSIS is performed a trend prediction through the combination of EMD and ELM. Every day one stock is selected for purchase and sold the next day.

The remainder of the paper is structured as follows. In the section \ref{sec-technical analysis} is described the technical indicators used in the work. In the section \ref{sec-methods} the techniques used in the work are presented. In the section \ref {sec-framework} a hybrid method for stock trading is proposed. In the section \ref{sec-results} the experimental results for the Brazilian stock market are presented and discussed, and in section \ref{sec-conclusions} the conclusions are drawn and directions for future work are indicated.

\section{Technical Analysis}
\label{sec-technical analysis}

The technical analysis consists of analyzing the behavior of a financial market in order to predict future price trends \cite{murphy1999}. The technical indicators are calculated using the historical series of prices and volume. There are several types of indicators, some of which are trend followers, price oscillators, volume analysis, among others.

\subsection{Relative Strength Index (RSI)}

RSI is an indicator formulated by Wilder \cite{wilder1978}. It measures the relationship between buying and selling forces in the market for a particular stock. It is calculated by:

\begin{equation}
\label{rsi}
RSI = 100-\frac{100}{1+RS}
\end{equation}
where $RS$ is the relative force, calculated by: 
\begin{equation}
\label{rs}
RS=\frac{mean(\Delta_{pos},n)}{mean(\Delta_{neg},n)}
\end{equation}
where $mean(\Delta_{pos},n)$ is the average of the positive changes (when the price of day $t$ is greater than the price of day $t-1$) for last $n$ periods, and $mean(\Delta_{neg},n)$ is the average of the negative changes (when the price of day $t$ is less than the price of day $t-1$) for last $n$ periods.

The indicator, in general, is used to define overbought and oversold areas. When the indicator is below of a lower limit (usually 30), it can be interpreted that the price of the asset is in an oversold zone, i.e, the selling force is losing strength and this may be a signal that the price will rise. Above of the upper limit (usually 70) the price is in the overbought zone indicating that the buying force is losing strength and the price is likely to fall.

The indicator can be used in different ways. Usually when the indicator crosses any of the limits, it is generated either a buy or sell signal. The trading rule for the RSI is described in Table \ref{tab-rsi-rule}.

\begin{table}[ht]
\centering
\caption{RSI rule}
\label{tab-rsi-rule}
\begin{tabular}{|l|l|}
\hline
Condition    & Signal \\\hline
if $RSI_{t-1}<30$ and $RSI_t>30$ & buy \\\hline
if $RSI_{t-1}>70$ and $RSI_t<70$ & sell \\\hline
otherwise & neutral \\\hline
\end{tabular}
\end{table}

If the value of the indicator crosses the lower limit from bottom to top, it is generated a buy signal. If the indicator crosses the upper limit from top to bottom, it is generated a sell signal. If there is no interception, no decision is made, i.e., the signal is neutral.

\subsection{Stochastic oscillator}

The stochastic oscillator is an indicator of momentum developed by George Lane \cite{Treuherz2008}. It can be interpreted similarly to the RSI, but the most used lower and upper limits are 20 and 80, respectively. This indicator is formed by two lines: $\%K$ and $\%D$. $\%K$ is calculated by:

\begin{equation}
\label{k}
\%K = \frac{C_{t}-min(L_{t},\dots,L_{t-n})}{max(H_{t},\dots,H_{t-n})-min(L_{t},\dots,L_{t-n})}\cdot 100
\end{equation}
where $C$ is the closing price of the day, $L$ is the minimum price of the day, $H$ is the maximum price of the day and $n$ is the number of periods used.

The line $\%D$ is a moving average of the line $\%K$, calculated by:
\begin{equation}
\label{d}
\%D = \frac{\sum_{i=0}^{n-1}\%K_{t-i}}{n}
\end{equation}

The trading rules for the $\%K$ and $\%D$ lines are similar to the RSI. The trading rule for the line $\%K$, called 'rule $\%K$' is described in Table \ref{tab-K-rule}. 

\begin{table}[ht]
\centering
\caption{ K rule}
\label{tab-K-rule}
\begin{tabular}{|l|l|}
\hline
Condition    & Signal \\\hline
if $\%K_{t-1}<20$ and $\%K_t>20$ & buy \\\hline
if $\%K_{t-1}>80$ and $\%K_t<80$ & sell \\\hline
otherwise & neutral \\\hline
\end{tabular}
\end{table}

The trading rule for the line $\%D$  called 'rule $\%D$' is described in Table \ref{tab-D-rule}.

\begin{table}[ht]
\centering
\caption{D rule}
\label{tab-D-rule}
\begin{tabular}{|l|l|}
\hline
Condition    & Signal \\\hline
if $\%D_{t-1}<20$ and $\%D_t>20$ & buy \\\hline
if $\%D_{t-1}>80$ and $\%D_t<80$ & sell \\\hline
otherwise & neutral \\\hline
\end{tabular}
\end{table}

Another possible interpretation for the indicator is at the intersections of the two lines. The rule for crossing the lines, called 'KD rule' is described in Table \ref{tab-KD-rule}.

\begin{table}[ht]
\centering
\caption{KD rule}
\label{tab-KD-rule}
\begin{tabular}{|l|l|}
\hline
Condition    & Signal \\\hline
if $\%K_{t-1}<\%D_{t-1}$ and $\%K_{t}>\%D_{t}$ & buy  \\\hline
if $\%K_{t-1}>\%D_{t-1}$ and $\%K_{t}<\%D_{t}$ & sell \\\hline
otherwise & neutral \\\hline
\end{tabular}
\end{table}

When the $\%K$ line crosses the $\%D$ line upwards is a buy signal; otherwise it is a sell signal. If there is no crossing of lines, the signal remains neutral.

\subsection{Commodities Channel Index (CCI)}

The CCI is an indicator developed by Donald R. Lambert, which compares the current price with a moving average of a given period and then normalizes the oscillator using a divisor based on the mean deviation \cite{murphy1999}. The indicator is calculated by:

\begin{equation}
\label{cci}
CCI= \frac{P-mean(P,n)}{0.015\cdot \sigma(P)}
\end{equation}
where $P$ is the typical price, and $\sigma(P)$ is the absolute mean deviation of the typical price.

The typical price is calculated by:

\begin{equation}
\label{preco_tipico}
P=\frac{C_{t}+L_{t}+H_{t}}{3}
\end{equation}
where $C_t$ is the closing price of the day, $L_t$ is the minimum price of the day and $H_t$ is the maximum price of the day.

The trading rule for the indicator, called the 'CCI rule', is describe in Table \ref{tab-CCI-rule}.

\begin{table}[ht]
\centering
\caption{CCI rule}
\label{tab-CCI-rule}
\begin{tabular}{|l|l|}
\hline
Condition    & Signal \\\hline
if ($CCI_{t-1}<-100$ and $CCI_t>-100$) or ($CCI_{t-1}<100$ and $CCI_t>100$) & buy \\\hline
if ($CCI_{t-1}>-100$ and $CCI_t<-100$) or ($CCI_{t-1}>100$ and $CCI_t<100$) & sell \\\hline
otherwise & neutral \\\hline
\end{tabular}
\end{table}

A buy signal is given when the indicator crosses the level of 100 or -100 from bottom to top. The sale signal occurs when the indicator crosses these same levels, but from top to bottom.

\section{Background on methods to select stocks}
\label{sec-methods}

\subsection{Technique for order preference by similarity to the ideal solution (TOPSIS)}
\label{sec-topsis}

The TOPSIS technique receives as input a matrix formed by the alternatives and criteria. Consider the decision matrix $A$:

\begin{equation}
A=\begin{bmatrix}
x_{11} & x_{12} & \ldots & x_{1n} \\ 
x_{21} & x_{22} & \ldots & x_{1n} \\ 
\vdots  & \vdots  & \ddots & \vdots \\ 
x_{m1} & \ldots  & \ldots & x_{mn} 
\end{bmatrix} 
\end{equation}
where each row is an alternative, each column is a criterion and $x_{ij}$ is the rating of the alternative $i$ according to criterion $j$.

As the criteria may not necessarily be of the same scale, they need to be normalized. The most well-known normalization methods are: vector normalization, transformation in linear scale as maximum-minimum, maximum, and sum \cite{ccelen2014}. The formulas are described by Equations \ref{linear}, \ref{max-min}, \ref{max} e \ref{sum}, respectively.

\begin{equation}
\label{linear}
r_{ij}=\frac{x_{ij}}{\sqrt{\sum_{i=1}^{m}x_{ij}^2}}
\end{equation}

\begin{equation}
\label{max-min}
r_{ij}=\frac{x_{ij}-x_{j}^{min}}{x_{j}^{max}-x_{j}^{min}}
\end{equation}

\begin{equation}
\label{max}
r_{ij}=\frac{x_{ij}}{x_{j}^{max}}
\end{equation}

\begin{equation}
\label{sum}
r_{ij}=\frac{x_{ij}}{\sum_{i=1}^{m}x_{ij}}
\end{equation}
where $r_{ij}$ is an element of the normalized matrix $R$.

Since some criteria may be more important than others for decision making, then they need to be weighted by $w_{j}$ weights. The weights must satisfy:

\begin{equation}
\label{w}
\sum_{j=1}^{n}w_{j}=1
\end{equation}

Next, it is necessary to weigh each column of the matrix by its criterion weight according to:

\begin{equation}
\label{v}
v_{ij}=r_{ij} \cdot w_{j} 
\end{equation}
where $v_{ij}$ is an element of the weighted matrix $V$.

Before describing the TOPSIS algorithm you need to define the type of each criterion. There are two types of criteria: cost and benefit. The cost criterion indicates that the lower the value, the better it is. Otherwise, the benefit criterion indicates that the higher the value, the better it is.

The first step of TOPSIS is to calculate the positive ideal solution and the negative ideal solution given by Equations \ref{fis}, \ref{nis}, \ref{vmax} and \ref{vmin}.

\begin{equation}
\label{fis}
v^{+}=(v_{1}^{+}, v_{2}^{+}, \dots, v_{n}^{+})
\end{equation}
\begin{equation}
\label{nis}
v^{-}=(v_{1}^{-}, v_{2}^{-}, \dots, v_{n}^{-})
\end{equation}
where:

\begin{equation}
\label{vmax}
    v_{j}^{+}=
    \left\{\begin{array}{rll}
    \max(v_{ij}), i=1,2, \dots, m  & \hbox{if criterium is benefit} \\
    \min(v_{ij}), i=1,2, \dots, m   & \hbox{if criterium is cost} 
\end{array}\right.
\end{equation}

\begin{equation}
\label{vmin}
    v_{j}^{-}=
    \left\{\begin{array}{rll}
    \min(v_{ij}), i=1,2, \dots, m  & \hbox{if criterium is benefit} \\
    \max(v_{ij}), i=1,2, \dots, m   & \hbox{if criterium is cost} 
\end{array}\right.
\end{equation}

The second step is to calculate the Euclidean distance of each alternative for each criterion in relation to the ideal positive solution and the ideal negative solution according to the Equations \ref{dist_pos} and \ref{dist_neg}.

\begin{equation}
\label{dist_pos}
d_{i}^{+}=\sqrt{\sum_{j=1}^{n}(v_{j}^{+}-v_{ij})^2}
\end{equation}

\begin{equation}
\label{dist_neg}
d_{i}^{-}=\sqrt{\sum_{j=1}^{n}(v_{j}^{-}-v_{ij})^2}
\end{equation}
where $d_{i}^{+}$ is the positive distance of alternative $i$ and $d_{i}^{-}$ is the negative distance of alternative $i$.

The third step is to calculate the relative closeness coefficient  $\xi_{i}$ for each alternative, as

\begin{equation}
\label{xi}
\xi_{i}=\frac{d_i^{-}}{d_i^{-}+d_i^{+}}
\end{equation}

Next, the alternatives are ranked in relation to their values of $\xi_{i}$; the higher the value, the better.

In this work, for the best alternative (in this case an stock) chosen, it is carried out a prediction of the tendency to increase the reliability of the selection. The empirical mode decomposition (EMD) is a promising technique for the extraction of the trend component.

\subsection{The Empirical mode decomposition (EMD)}
\label{sec-emd}

Empirical mode decomposition is a technique of data analysis proposed by Huang et al. \cite{huang1998}. The decomposition is based on the direct energy extraction associated with several intrinsic time scales \cite{huang1998}. Each sub-series generated by the decomposition is called the intrinsic mode function (IMF), where the last one, is called residuum. An IMF is a function that must satisfy two conditions: $(1)$ for any data set, the number of extremes and the number of crosses at zero must be equal or  to differ by a maximum of one; and $(2)$ at any point, the average value of the envelope defined by local minima and maxima should be zero \cite{huang1998}.

The method of decomposition consists firstly in identifying all local maxima and minima of a time series $X$. After this, the local maxima points are connected by means of a cubic interpolation forming the upper envelope $u$. Similarly, By connecting the local minima is formed the lower envelope $l$. A new time series $m_1$ is formed by the mean of the two envelopes:  $(u+l)/2$. Subtracting this series of the original one, it results of the first component ($h_1$) according to:

\begin{equation}
    h_1= X - m_1
\end{equation}

However, this component needs to satisfy both IMF conditions to become one. If this fact does not occur the process is done again, but this time, the series $h_1$ is treated as the original series. In this way, the envelopes are calculated from $h_1$, and the mean is calculated ($m_{11}$). A new component $h_{11}$ is then calculated as:

\begin{equation}
    h_{11}= h_1 - m_{11}
\end{equation}

Again, it is checked whether this component meets the conditions of an IMF. If it does not satisfy, the process is repeated again. This process is repeated until an IMF is obtained:

\begin{equation}
    h_{1k}= h_{1(k-1)} - m_{1k}
\end{equation}

So, $h_{1k}$ becomes the first IMF ($c_1$)) of the series:

\begin{equation}
    c_1=h_{1k}
\end{equation}

The remainder of the series is called residuum ($r$), calculated as:

\begin{equation}
    r_1=X-c_1
\end{equation}

From the residuum generated by the subtraction of the first component ($r_1$), is calculated the second component ($c_2$), repeating the same procedure. The process is repeated $n$ times, until no further IMFs can be extracted. So, the series can be reconstructed by:

\begin{equation}
    X= \sum_{i=1}^{n}c_i + r_n 
\end{equation}
where $n$ is the number of IMFs generated and $r_n$  is the final residuum of the series.

\begin{figure}[ht]
    \centering
    \includegraphics[width=12cm, height=2cm]{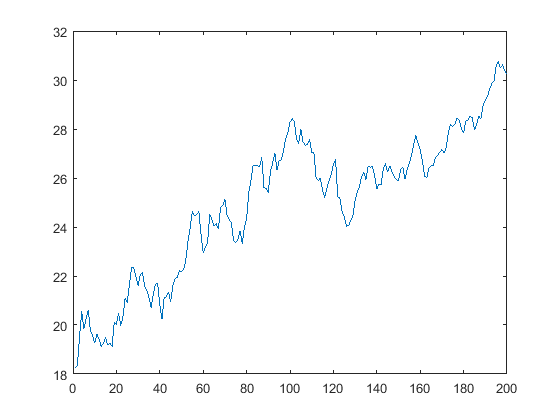}
    \includegraphics[width=12cm, height=2cm]{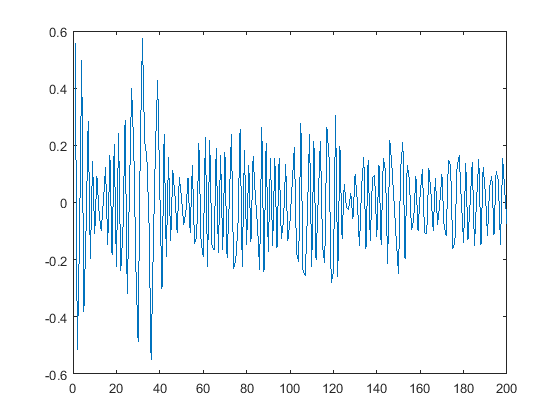}
    \includegraphics[width=12cm, height=2cm]{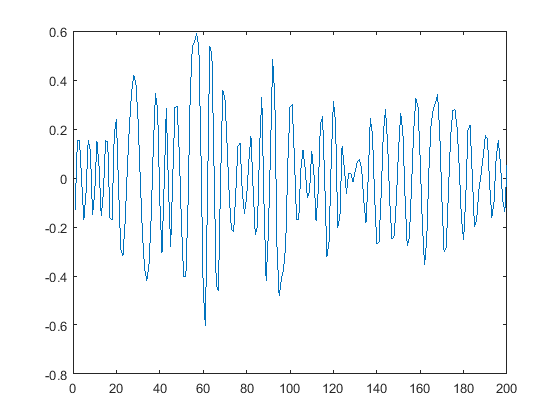}
    \includegraphics[width=12cm, height=2cm]{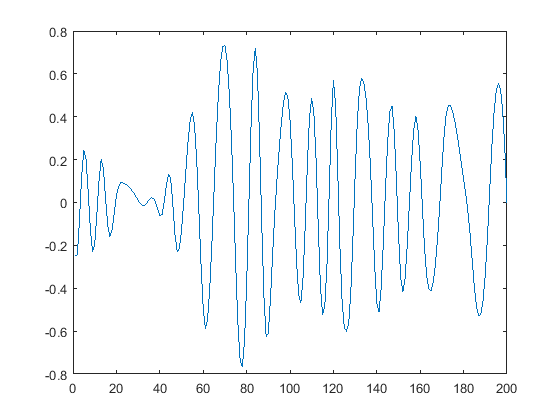}
    \includegraphics[width=12cm, height=2cm]{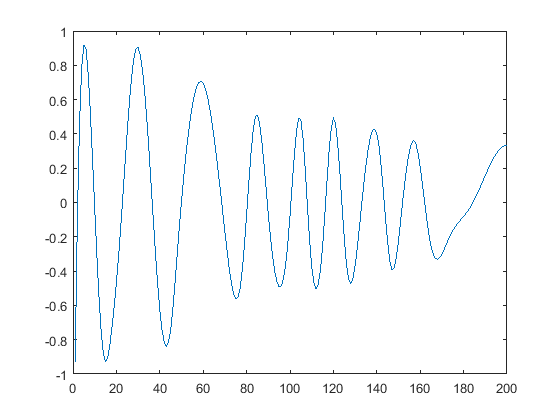}
    \includegraphics[width=12cm, height=2cm]{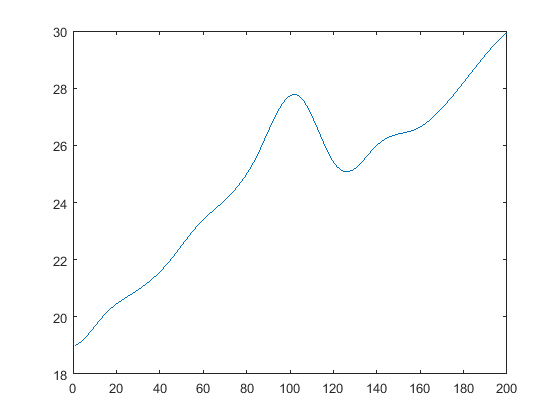}
    \caption{ Time series decomposition performed by EMD}
    \label{fig:imfs}
\end{figure}

Figure \ref{fig:imfs}  illustrates the decomposition of a time series carried out by EMD. At the top of Figure \ref{fig:imfs} is the original serie, which is followed by the first, second, third and fourth IMF and finally the residuum.

Since there are several components of the decomposed series, an artificial neural network will be used to predict the trend component (in this case, the series residuum).

\subsection{Extreme Learning Machine (ELM)}
\label{sec-elm}

An artificial neural network (ANN) is a parallel and distributed information processing structure that consists of processing elements interconnected through unidirectional signals \cite{nielsen1988}. ANNs are formed by the connection of several processing elements, called artificial neurons, each one producing a sequence of activations of real values \cite{schmidhuber2015}.

\begin{figure}[ht]
    \centering
    \includegraphics[width=1\textwidth]{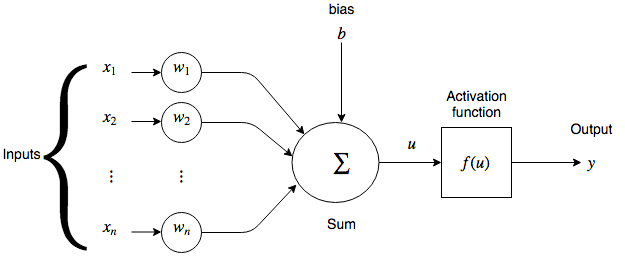}
    \caption{Artificial neuron}
    \label{fig:neuronio}
\end{figure}

The artificial neuron is the basic unit of processing of neural networks. It receives input signals $x_i$, which propagate through channels called connections. Each connection has a certain value, called weight $w_i$, and it weights the inputs ($x_i \cdot w_i$). The weighted inputs are summed and added to this sum is a bias ($b$), generating a single signal ($u$). This signal passes through an activation function $f(u)$, resulting in the output of the neuron ($y$). Figure \ref{fig:neuronio} illustrates an artificial neuron. Examples of commonly used activation functions are: logistic sigmoidal, hyperbolic tangent and rectified linear unit (ReLu).

\begin{figure}[h]
    \centering
    \includegraphics[width=0.7\textwidth]{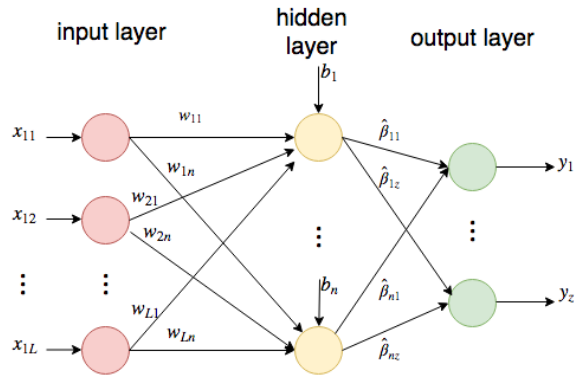}
    \caption{Architecture of a single layer feedforward network (SLFN)}
    \label{fig:arquitetura}
\end{figure}

Neurons connect to each other and usually they are grouped in layers. The first layer is the input layer, where no processing takes place, only the transmission of the input signals. The following layers, except the last layer, are called hidden layers, which contain one or more layers and are responsible for signal processing. The last layer aims to present the network response to the problem at hand. Figure \ref{fig:arquitetura} shows the architecture of a neural network with a single hidden layer.

In ANNs with one hidden layer the main difference for other neural networks is that training occurs in a deterministic form. The weights between the initial layer and the hidden layer, as well as the bias, are initialized at random. The weights between the hidden layer and the output layer are determined analytically. 

The extreme learning machine (ELM) is a learning algorithm for artificial neural networks with a single hidden layer \cite{huang2004}. The first step of the ELM algorithm is the construction of the $H$ matrix, which is the matrix resulting from the neuron output of the hidden layer for each of the input samples according to:

\begin{equation}
\label{matriz_h}
H = \begin{bmatrix}
f(x_{1}w_{1}+b_{i}) & \dots  & f(x_{1}w_{n}+b_{m}) \\
\vdots  & \ddots & \vdots \\
f(x_{m}w_{1}+b_{i}) & \dots  & f(x_{m}w_{n}+b_{n})
\end{bmatrix}
\end{equation}
where each row corresponds to one sample and each column represents a hidden layer neuron,  $m$ is the number of input samples, and  $n$ is the number of neurons in the hidden layer. In this case, $x_i$ is a sample composed of vector of attributes and $w_i$ is the weights vector of a neuron.

The second step is to calculate the weight matrix between the hidden layer and the output layer, called $\hat{\beta}$. It is calculated as the least squares solution for the linear system $H\hat{\beta}=T$, with $\hat{\beta}=H^{\dagger}T$ where $H^{\dagger}$ is the generalized Moore-Penrose inverse of the $H$  matrix and $T$ is the response vector for the input samples \cite{huang2004, huang2006}.

The pseudo-code of the ELM training is presented in algorithm 1.

 \begin{algorithm}
 \caption{ELM training}
 \begin{algorithmic}[1]
 \STATE \textbf{Initialization}:
 \STATE \ \ \ Preprocessing of the input samples
 \STATE \ \ \ Initialization of weights and bias between the first and hidden layer with 
 \ \ \ random values
 \STATE \ \ \ Set the number of neurons in the hidden layer
 \STATE Calculation of the matrix $H$ according \ref{matriz_h}
 \STATE Calculation of $\hat{\beta}$ according $\hat{\beta}=H^{\dagger}T$
 \end{algorithmic} 
 \end{algorithm}


\section{Development of a hybrid method for stock trading}
\label{sec-framework}

\subsection{Introduction}

In order to decide which stock to buy and at what time it should be bought, a new methodology was developed combining several techniques. In the first part, the selection of the stock is made by TOPSIS, which selects the most suitable stock for purchase. In this set, only stocks with a large volume and that belong of the Bovespa index are considered. So, every day a stock is bought and sold in the next day. In the second part, the stock selected by TOPSIS goes through a confirmation for the purchase. It consists of a  hybrid model made up of EMD and ELM. First the time series of the stock is decomposed and the trend is extracted and in turn ELM classifies the stock price trend as high or low. If the trend is high, then the purchase is performed, otherwise it is declined.

\subsection{Selection of stock for purchase through TOPSIS}

\begin{figure}[ht]
    \centering
    \includegraphics[width=0.5\textwidth]{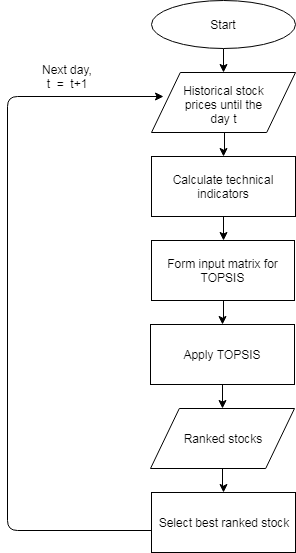}
    \caption{TOPSIS using technical analysis indicators}
    \label{fig:topsis}
\end{figure}

For stock selection, historical values of stock prices and volume are first collected and from these data, the technical indicators for each stock on day $t$ are calculated. Next, the input matrix is formed with the alternatives and criteria for the decision-making process. The alternatives are the stocks, and the criteria are the technical indicators. So, the input matrix is made up of rows (stock number), and columns (number of criteria). The criteria used were: the 14-day RSI indicator, stochastic $\%$K line (14 days), stochastic $\%$D (3-days), and the CCI (14 days). They all are cost criteria. The weight of each criterion was the same for all criteria. The TOPSIS is applied to the matrix in order to perform the ranking of stocks. The best ranked stock is selected and advances to the next day ($t+1$). Then,  it is checked whether its price has increased or decreased and the value of that variation. Thereafter, the process is repeated several times. It is important to note that selecting an stock with TOPSIS also means dealing with the dynamic and stochastic behavior of financial time series. 
The methodology of selection of stocks based on TOPSIS using criteria of technical analysis is illustrated in Figure \ref{fig:topsis}.

For trading purposes the stock is selected at the end of the day, i.e. it is bought at the closing price. The sale of the stock takes place at the end of the next day and another stock is then bought. For instance, let us suppose that TOPSIS selects an stock in a set of ten stocks on a given day. The stock data are collected and the technical indicators are calculated building the decision matrix as shown in Table \ref{tab-exemplo-topsis-1}.

\begin{table}[ht]
\small
\centering
\caption{Example illustrating the application of TOPSIS for stock selection}
\label{tab-exemplo-topsis-1}
\begin{tabular}{|l|l|l|l|l|l|l|}
\hline
Alternative /criteria & RSI & stochastic K & stochastic D & CCI \\ \hline
Stock A & 37.46 &  7.43 &  9.13 &  -86.35 \\ \hline
Stock B & 7.58 &  3.48 &  2.33 &  -166.79 \\ \hline
Stock C & 17.65 &  5.41 &  4.27 &  -150.15 \\ \hline
Stock D & 29.47 &  14.29 &  13.08 &  -105.43 \\ \hline
Stock E & 49.13 &  10.80 &  19.73 &  -65.99 \\ \hline
Stock F & 20.93 &  8.62 &  6.12 &  -97.30 \\ \hline
Stock G & 14.00 &  4.80 &  4.12 &  -106.82 \\ \hline
Stock H & 17.71 &  1.09 &  2.71 &  -194.24 \\ \hline
Stock I & 26.38 &  4.52 &  6.57 &  -98.28 \\ \hline
Stock J & 20.16 &  4.16 &  3.40 &  -154.57 \\ \hline
\end{tabular}
\end{table}

Based on the data provided TOPSIS performs the ranking of stocks according to Table \ref{tab-exemplo-topsis-2}. This way, at the end of the first day, the 'H' stock will be purchased. At the end of the next day, it will be sold and based on new data TOPSIS will select another stock for purchase and the process is repeated every day.

\begin{table}[ht]
\small
\centering
\caption{Results of the application of TOPSIS for stock selection}
\label{tab-exemplo-topsis-2}
\begin{tabular}{|l|l|l|}
\hline
Alternative & Value of $\xi$ & Position\\ \hline
Stock A & 0.4049 & 8\\ \hline
Stock B & 0.8660 & 2\\ \hline
Stock C & 0.7336 & 4\\ \hline
Stock D & 0.3179 & 9\\ \hline
Stock E & 0.1231 & 10\\ \hline
Stock F & 0.5285 & 7\\ \hline
Stock G & 0.6578 & 5\\ \hline
Stock H & 0.8847 & 1\\ \hline
Stock I & 0.5632 & 6\\ \hline
Stock J & 0.7590 & 3\\ \hline
\end{tabular}
\end{table}

\subsection{Trend Classification by using EMD-ELM}

The EMD-ELM hybrid model consists of the decomposition of the price series by EMD and the prediction of the price trend by the ELM neural network. In this work, only the main component extracted by the EMD is used as input to the ELM. The first components extracted by the EMD are of higher frequencies and also may contain noise. In this paper, it is proposed to use only the last component, the residuum. Usually, for financial series, this component contains the highest percentage of the price value. In addition, it contains the price trend.

After selecting the best stock at the end of the day, the historical price series of that stock is used for EMD application. $p$ is used for previous days, with the last element of the series being the current day $t$. In this way, the series $S=[x_{t-p},x_{t-p+1}, \dots, x_t]$, where $x_i$ is the stock price on a given day. The series is then normalized between 0 and 1, according to:

\begin{equation}
    S_i = \frac{S_i- min(S)}{max(S)-min(S)}
\end{equation}

By applying the EMD to this series, several IMFs are generated, the last is the residuum, which is used by the ELM. The series of residues $[r_1,r_2,\cdots,r_n]$ is modeled to be used as input to the neural network according to Table \ref{tab:modelagem}.

\begin{table*}[ht]
\caption{Time Series Modeling}
\label{tab:modelagem}
\centering
\begin{tabular}{|cccc|c|}
\hline
\multicolumn{4}{|l|}{inputs} & target \\ \hline
$r_{1}$ & $r_{2}$ & \dots & $r_{1+(\mu-1)}$    & $r_{1+\mu}$\\
\hline
$r_{2}$ & $r_{3}$ & \dots & $r_{2+(\mu-1)}$    & $r_{2+\mu}$\\
\hline
 $\vdots$ & $\vdots$ & $\ddots$ & $\vdots$ & $\vdots$ \\
\hline
$r_{n-\mu}$ & $r_{n-(\mu-1)}$ & \dots & $r_{n-1}$ & $r_{n}$\\
\hline
\end{tabular}
\end{table*}

The time series data are processed in $\mu$ columns (number of delays) representing the network inputs and a column representing the expected values (target) for the output of the neural network. The neural network is then trained from the input matrix. The weights are generated randomly in the range of -1 to 1, and the activation function used was the sigmoidal. So, according to the ELM algorithm, the matrix $\hat{\beta}$ is calculated. For a neural network already trained, the test data are presented to it: $[r_{n-(\mu-1)}, r_{n-(\mu-2)}  \dots  r_{n}]$. The network response is the expected value of  $r_{n+1}$. The real value of $r_{n+1}$ is not known as it is in the future. Also, it can not be known in the future, because by including the next day price value, the series structure changes and EMD is applied again, and different IMFs are generated. However, in our case the goal is to find out if the trend is high or low. This is done by comparing the last value of the component ($r_n$) with the output of the neural network ($y$), which is the prediction for $r_{n+1}$, as given by Equation \ref{eq:trend}:

 \begin{equation}
\label{eq:trend}
    trend=
    \left\{\begin{array}{rll}
    high  & \hbox{if $y > r_n$} \\
    low & \hbox{if $y \leq r_n$} 
\end{array}\right.
\end{equation}

\begin{figure}[ht]
    \centering
    \includegraphics[width=0.9\textwidth]{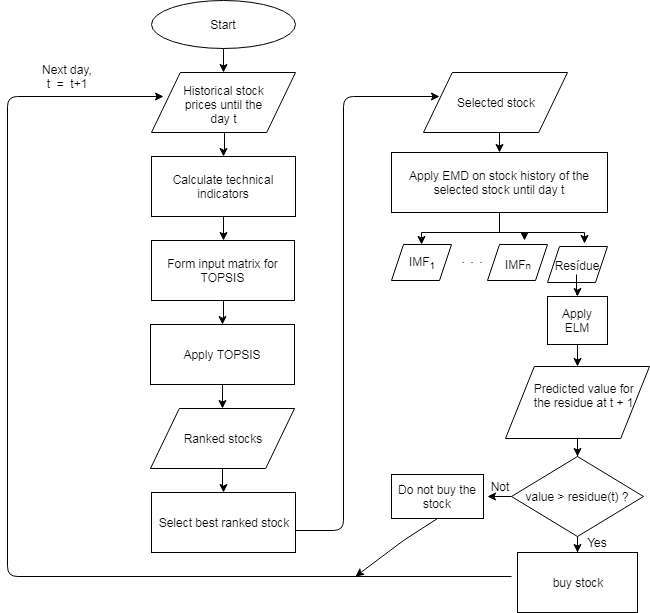}
    \caption{Method combining TOPSIS with EMD-ELM}
    \label{fig:topsis-emd-elm}
\end{figure}

If the neural network indicates an increase in value, then the purchase of the stock is confirmed, otherwise no stock is purchased. Of course, this reduces the total number of trades. The method combining TOPSIS and EMD-ELM is illustrated in Figure \ref{fig:topsis-emd-elm}.

Suppose that TOPSIS has selected a particular stock and that a window of 20 past prices (a  total amount of 21 with the price of the day) is used, as shown in Table  \ref{tab:serie-exemplo1}. The series is first normalized and then decomposed by the EMD, as in Table  \ref{tab:serie-exemplo2}. Three components are generated after applying EMD. The last component is modeled as input to the neural network. Modeling using 6 delays is shown in Table \ref{tab:serie-exemplo3}. Based on these data, the neural network using  ELM is trained. Then, the test data for ELM is presented: [0.5720, 0.5844, 0.5952, 0.6036, 0.6096, 0.6134].

\begin{table}[ht]
\caption{Time series of one stock selected by TOPSIS}
\label{tab:serie-exemplo1}
\centering
\begin{tabular}{|cc|cc|cc|}
\hline
Time & Price & Time & Price & Time & Price\\ \hline
$t-20$ &  11.23   & $t-13$ & 15.87 & $t-6$ &    12.92\\
$t-19$ &    10.57 & $t-12$ & 16.85 & $t-5$ &    14.21\\
$t-18$ &    10.55 & $t-11$ & 14.41 & $t-4$ &    14.86\\
$t-17$ &    11.32 & $t-10$ & 13.90 & $t-3$ &    14.70\\
$t-16$ &    12.28 & $t-9$ & 13.42  & $t-2$ &    14.52\\
$t-15$ &    13.63 & $t-8$ & 13.30  & $t-1$ &    14.79\\
$t-14$ &    14.98 & $t-7$ & 12.89  & $t$   &    13.54\\ 

\hline
\end{tabular}
\end{table}

\begin{table}[ht]
\caption{Application of the EMD on the normalized series}
\label{tab:serie-exemplo2}
\small
\centering
\begin{tabular}{|c|c|c|c|c|}
\hline
Time & Normalized price & IMF1 & IMF2 & Residuum\\ \hline
$t-20$ &0.1079 &  -0.4825 &  -0.0694 &  0.6599\\
$t-19$ &0.0032 &  -0.5781 &  -0.0576 &  0.6388\\
$t-18$ &0      &  -0.5786 &  -0.0380 &  0.6167\\
$t-17$ &0.1222 &  -0.4592 &  -0.0130 &  0.5944\\
$t-16$ &0.2746 &  -0.3122 &  0.0136  &  0.5733\\
$t-15$ &0.4889 &  -0.1029 &  0.0376  &  0.5543\\
$t-14$ &0.7032 &  0.1099  &  0.0547  &  0.5386\\
$t-13$ &0.8444 &  0.2563  &  0.0609  &  0.5272\\
$t-12$ &1.0000 &  0.4267  &  0.0525  &  0.5208\\
$t-11$ &0.6127 &  0.0652  &  0.0285  &  0.5190\\
$t-10$ &0.5317 &  0.0150  &  -0.0044 &  0.5212\\
$t-9$ &0.4556 &  -0.0336 &  -0.0378 &  0.5269\\
$t-8$ &0.4365 &  -0.0359 &  -0.0632 &  0.5356\\
$t-7$ &0.3714 &  -0.1031 &  -0.0721 &  0.5466\\
$t-6$ &0.3762 &  -0.1256 &  -0.0572 &  0.5591\\
$t-5$ &0.5810 &  0.0280  &  -0.0190 &  0.5720\\
$t-4$ &0.6841 &  0.0677  &  0.0320  &  0.5844\\
$t-3$ &0.6587 &  -0.0145 &  0.0780  &  0.5952\\
$t-2$ &0.6302 &  -0.0524 &  0.0789  &  0.6036\\
$t-1$ &0.6730 &  0.0555  &  0.0080  &  0.6096\\
$t$ &0.4746 &  -0.0590 &  -0.0798 &  0.6134\\
\hline
\end{tabular}
\end{table}

\begin{table*}[ht]
\caption{Series modeled as input for neural network}
\label{tab:serie-exemplo3}
\centering
\begin{tabular}{|cccccc|c|}
\hline
\multicolumn{6}{|l|}{inputs} & target \\ \hline
0.6599 &  0.6388 &  0.6167 &  0.5944 &  0.5733 &  0.5543 &  0.5386\\
0.6388 &  0.6167 &  0.5944 &  0.5733 &  0.5543 &  0.5386 &  0.5272\\
0.6167 &  0.5944 &  0.5733 &  0.5543 &  0.5386 &  0.5272 &  0.5208\\
0.5944 &  0.5733 &  0.5543 &  0.5386 &  0.5272 &  0.5208 &  0.5190\\
0.5733 &  0.5543 &  0.5386 &  0.5272 &  0.5208 &  0.5190 &  0.5212\\
0.5543 &  0.5386 &  0.5272 &  0.5208 &  0.5190 &  0.5212 &  0.5269\\
$\vdots$ &  $\vdots$ &  $\vdots$ &  $\vdots$ &  $\vdots$ &  $\vdots$ &  $\vdots$\\
0.5212 &  0.5269 &  0.5356 &  0.5466 &  0.5591 &  0.5720 &  0.5844\\
0.5269 &  0.5356 &  0.5466 &  0.5591 &  0.5720 &  0.5844 &  0.5952\\
0.5356 &  0.5466 &  0.5591 &  0.5720 &  0.5844 &  0.5952 &  0.6036\\
0.5466 &  0.5591 &  0.5720 &  0.5844 &  0.5952 &  0.6036 &  0.6096\\
0.5591 &  0.5720 &  0.5844 &  0.5952 &  0.6036 &  0.6096 &  0.6134\\
\hline
\end{tabular}
\end{table*}

The neural network generates an output of 0.6157 for example. Since 0.6157 is greater than the last value (0.6134), then the prediction of the neural network is a \textit{high} value of the component. So, the purchase of the stock is run at the current price (R\$13,54).

\subsection{Parameter Selection}

The hybrid method has several parameters that need to be set up. So, the settings  of the parameters is based on the results of the previous year and on two criteria: percentage of operations with profit and higher return. For TOPSIS, the technical indicators are used as criteria. In this work, RSI, stochastic K and D, and CCI. First they are used all together, and in turn combinations of two and three indicators will be tested. Based on the results of TOPSIS, the configuration with the highest percentage of profit and the one with the highest return are used for selection.  Then the EMD-ELM stock purchase confirmation methodology is integrated. For EMD-ELM, the most influential parameter is the size of the data sliding window (delay) used. It was tested for values in the range from 20 to 400, with increasing step of 10.

\subsection{Look-ahead effect}

The \textit{look-ahead bias} effect involves the use of information that would not yet be available at that time \cite{mahfoud1996}. That is, any form of future data usage can be characterized as \textit{look-ahead bias}. Even normalization techniques such as min-max and z-score, when applied over the whole time series are considered \textit{look-ahead bias} \cite{furlaneto2017}.

Many papers apply pre-processing or normalization techniques over the entire time series including the test data. This, as presented in \cite{furlaneto2017}, logically causes an improvement in results. However, for real-time applications such as stock trading such methodologies are not applicable since there is no way to know the coming data in future.

In this work, although the experiments are done with data from the past, a protocol similar to \cite{furlaneto2017} whereas data not yet available is not used. The hybrid method makes its decisions at the end of the day, only with the information available until that day. At no time the data are normalized based on data not yet available. The EMD technique is applied only to the prices available so far.

\section{Experimental results}
\label{sec-results}

\subsection{Dataset}

The database consists of the daily values of open, low, high and close prices and trade volume of several stocks of the Brazilian stock exchange. Data from the years 2016 and 2017 were collected using MetaTrader 5 software \cite{MetaTrader}. 50 stocks were used  by TOPSIS. Only stocks with a large volume of trading were considered, which are generally part of the Bovespa index. The stocks name  and their trading codes are shown in \ref{appendix-a}.

\subsection{Evaluation metrics}

The first metric used to evaluate the performance of the methodology was the accuracy of the selection. It measures the percentage of stock selections that had an increase in price the next day. The accuracy is calculated according to:

\begin{equation}
    Accuracy= \frac{\textit{number of correct stock selections}}{\textit{total number of stocks}}\cdot 100
\end{equation}

The other metric used was the cumulative return ($Rc$), which is basically the return in the investment period. It is calculated by:

\begin{equation}
    Rc=\prod_{i=1}^n(1+ \frac{C_{i+1}-C_i}{C_i})\end{equation}
where $C_i$ is the close price of one stock on day $i$ and $n$ is the total trading days.

The cumulative return describes the relation between the initial investment and the resulting final amount. For example a  $Rc$ of 2.5 means that when investing 1000 Dollars, the amount at the end of the investment is 2500 Dollars. However, in this study this value will be converted to percentage return, i.e., the percentage of profit generated calculated according to:

\begin{equation}
    \text{percentage \ return} = (Rc - 1)\cdot{100}
\end{equation}

The percentage return used in this study corresponds to the gross profit, that is, transaction costs and taxes are not considered.

\subsection{Results using TOPSIS}

\subsubsection{Comparison procedure}

To evaluate the selection of stocks made by TOPSIS, a random selection strategy is used for comparison. In this strategy, an stock is randomly selected every day. To generate statistical measures such as mean and standard deviation, the strategy was executed 1000 times.

The rate of return generated by the selection made by TOPSIS is compared to the return of the Bovespa index in the same trading period. The Bovespa index consists of a theoretical portfolio of the most traded stocks in the Brazilian market and has the objective of representing the average performance of the most representative stocks of the market \cite{bmfbovespa}. The random strategy was also used to compare in terms of percentage return.

\subsubsection{Application to the year 2016}

One of the most important steps in the use of TOPSIS is the choice of criteria, since they strongly influence the final result of the algorithm. All combinations of the indicators used were tested and the results were compared with the Bovespa index and the random strategy in terms of accuracy of selection and return percentage as shown in Table \ref{tab-2016}.

\begin{table}[ht]
\centering
\caption{Results of the application of TOPSIS for the year 2016}
\label{tab-2016}
\begin{tabular}{|l|l|l|l|}
\hline
Method    & Criteria & Accuracy  & Percentage return \\\hline
TOPSIS& RSI,    stoch. K                 &49.39\% & 66.48\%\\\hline
TOPSIS& RSI,    stoch. D                 &43.72\% & 14.36\%\\\hline
TOPSIS& RSI,    CCI                    &53.85\% & 134.84\%\\\hline
TOPSIS& stoch. K, stoch. D                 &54.25\% & 141.61\%\\\hline
TOPSIS& stoch. K, CCI                    &57.89\% & 248.39\%\\\hline
TOPSIS& stoch. D, CCI                    &57.89\% & \textbf{326.77\%}\\\hline
TOPSIS& RSI,    stoch. K, stoch. D         &49.39\% & 50.16\%\\\hline
TOPSIS& RSI,    stoch. K, CCI            &55.87\% & 222.13\%\\\hline
TOPSIS& RSI,    stoch. D, CCI            &52.63\% & 89.06\%\\\hline
TOPSIS& stoch. K, stoch. D, CCI            &\textbf{58.30\%} & 269.73\%\\\hline
TOPSIS & RSI,    stoch. K, stoch. D, CCI    &53.44\% & 133.70\%\\\hline
Random & - & $51.86\pm2.73\%$ & $55.15\pm59.94\%$\\\hline  
Ibovespa  & - & -              & 42.91\%\\ \hline
\end{tabular}
\end{table}

\subsubsection{Analysis of the criteria used by TOPSIS}
\begin{figure}[ht]
    \centering
    \includegraphics[width=13cm, height=3cm]{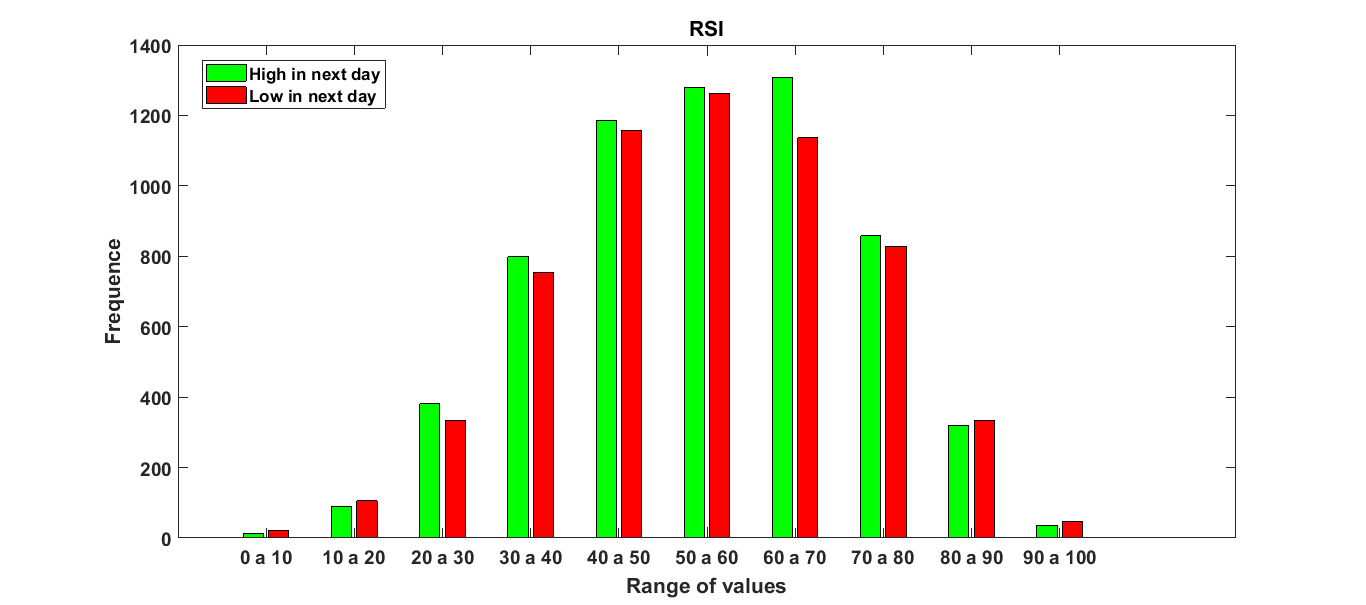}
    \includegraphics[width=13cm, height=3cm]{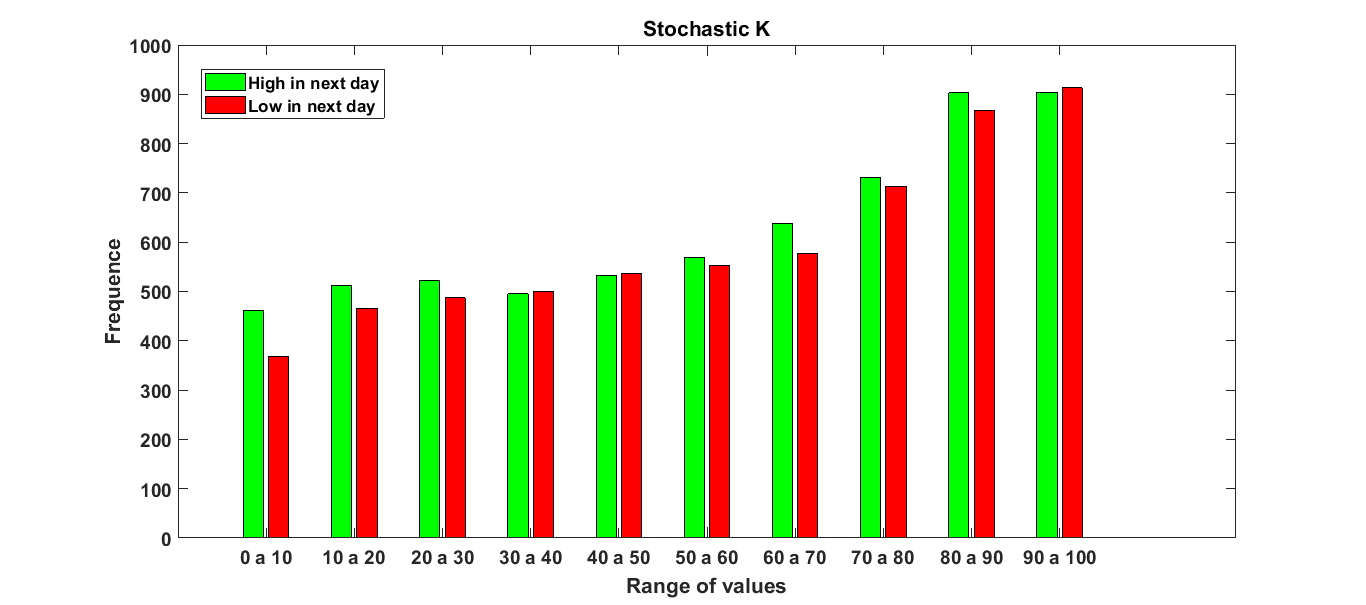}
    \includegraphics[width=12.5cm, height=3cm]{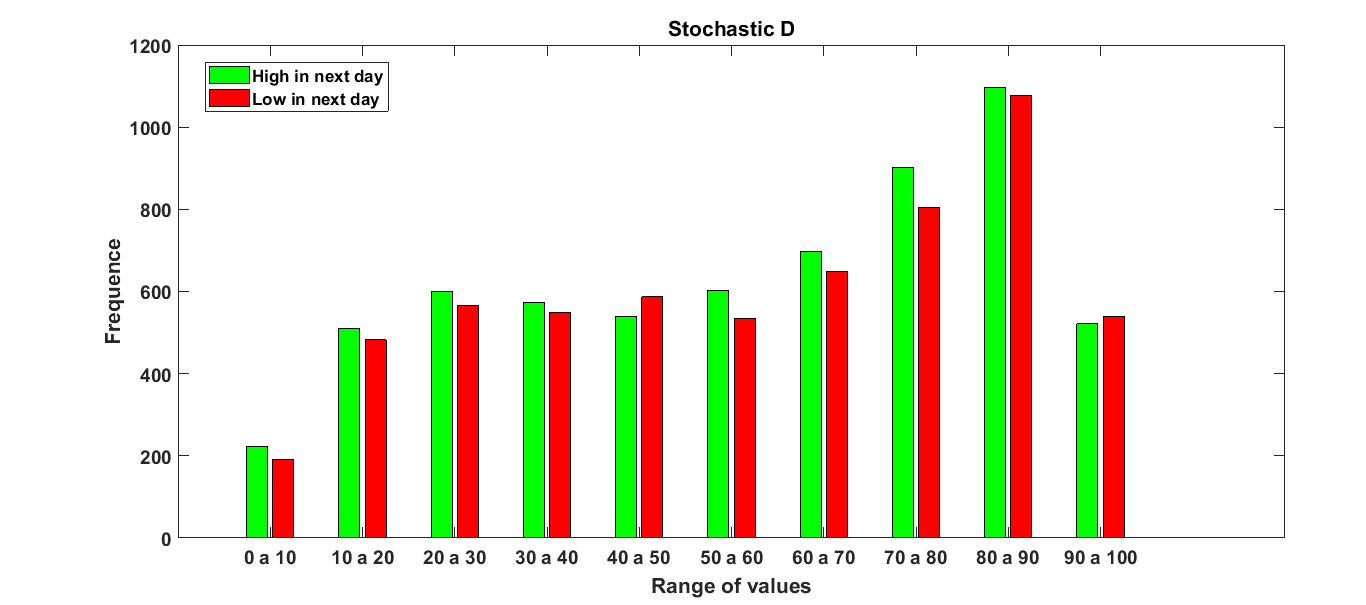}
    \includegraphics[width=13cm, height=3cm]{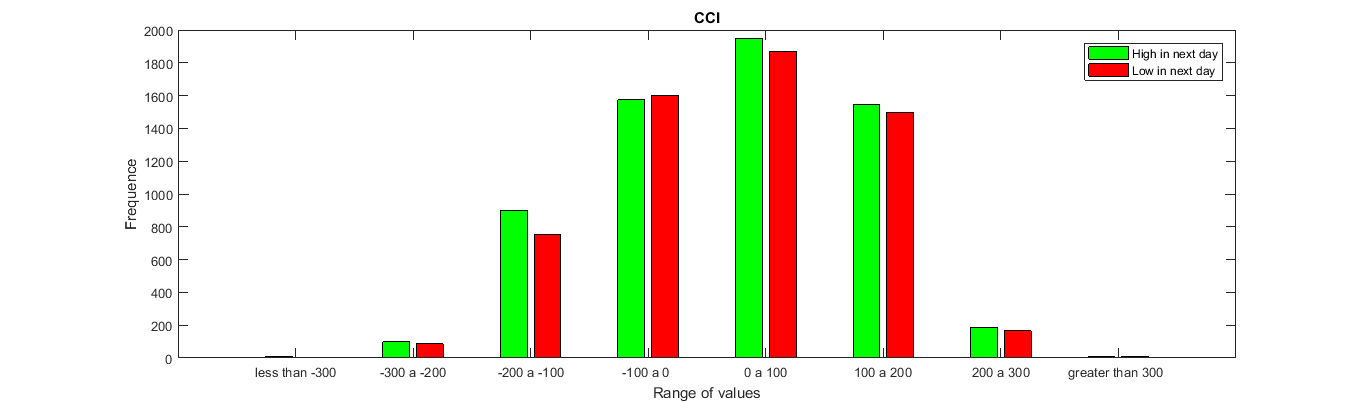}

    \caption{Histograms for the technical indicators}
    \label{fig:hist}
\end{figure}

As expected, the use of different criteria changes the results of TOPSIS. However, it is possible to observe that some indicators positively impact the results and others negatively. The CCI indicator is present in the best results both in accuracy and in percentage return. When using all indicators, for example, the accuracy is $53,44\%$ and the return is $133,70\%$. When the CCI indicator is removed, the accuracy drops to $49,39\%$ and the return to $50,16\%$. On the other hand, the RSI indicator is present in the worst results of the algorithm. By removing only this indicator, the accuracy rises to $58,30\%$ and the return to $269,73\%$. Indicators do not always behave according to their type of criterion. For instance, when a cost indicator presents a low value, the stock should rise the next day. However, the maximization or minimization of indicators is not always directly related to future stock price behaviour. This fact is illustrated by histograms in figure \ref{fig:hist}.
 
Histograms were constructed based on the values of the indicators for all 50 stocks. On the horizontal axis are the value ranges for the indicators and on the vertical axis the frequency, i.e., the number of times they occurred. For each value range there are two classes: high (represented by the green bar) and low (represented by the red bar). The 'high' class represents the cases in which the stock rises next day. For instance, one stock had the closing price of R\$ 20,00 (Brazilian currency) and the RSI value was 25 on day $t$ and the next day the closure price rises to R\$21,00. In this case, the RSI is counted for the 'high' class in the third value range (20 to 30) of the indicator histogram. 

In the RSI histogram it is possible to observe that when the value of the indicator is below 20 (first two range), the number of occurrences of the 'high' and 'low' classes are close, with a slight advantage to the latter. The same is true for the last two value ranges. The ideal case for TOPSIS would be that in the first classes where there were a dominance of the 'high' class.  In the histogram of stochastic K, one can notice a certain dominance of the 'high' class for the first three ranges. This shows that the indicator is able in a certain way to fit as a cost criterion. The stochastic indicator D presents a histogram similar to stochastic K. In the first four ranges, there is a slightly higher number of occurrences of the 'high' class. Like the previous indicator, it presents a certain adequacy as cost criterion. The CCI indicator histogram shows significance only from the second to the seventh value ranges. In the initial ranges (second and third),  the 'high' class is more frequent. So, it is to some extent compatible with the cost criterion.

The first part of the method acts by selecting the best stock  among the 50 available. For this, the TOPSIS is used with technical analysis indicators as criteria. The best results were the stochastic K and D and CCI (best accuracy) and the use of stochastic D and CCI (best return).

\subsection{Results using TOPSIS + EMD-ELM}

The second part of the method acts by confirming or not the purchase of the stock selected by TOPSIS. First, some parameters of the EMD-ELM have been setup by means of a empirical sensitivity study, as the amount of samples used for training each day (size of the training window), the number of delays used as input to ELM, and the number of neurons in the  hidden layer of ELM. The number of delays and the number of neurons were empirically defined as 6 and 20 respectively, since it was found through simulations experiments that these parameters did not present a great influence on the results. The parameter of greatest influence is the size of the training window. It was tested in the range of 20 to 400 varying the value from 10 with increment of 10. The tests were performed using TOPSIS with the two sets of best-performing criteria for the year 2016, described previously, and a third set using all indicators. The results using TOPSIS with criteria stochastic K and D and CCI for stock selection are shown in Table \ref{tab-result-topsis-emd-elm1}.

\renewcommand{\arraystretch}{0.8}
\begin{table}[ht]
\tiny
\centering
\caption{Results of TOPSIS (K-D-CCI) + EMD-ELM for the year 2016}
\label{tab-result-topsis-emd-elm1}
\begin{tabular}{|c|c|c|c|c|c|c|}
\hline
Window & Delay number & Neurons & Negotiations & Neg. with profit & Accuracy & Percentage return\\\hline
20 & 6 & 20 & 78 & 44 & 56.41 &  6.78 \\ \hline
30 & 6 & 20 & 70 & 38 & 54.29 &  0.23 \\ \hline
40 & 6 & 20 & 78 & 42 & 53.85 &  44.13 \\ \hline
50 & 6 & 20 & 68 & 43 & 63.24 &  78.68 \\ \hline
60 & 6 & 20 & 71 & 43 & 60.56 &  49.63 \\ \hline
70 & 6 & 20 & 75 & 47 & 62.67 &  57.95 \\ \hline
80 & 6 & 20 & 86 & 54 & 62.79 &  73.19 \\ \hline
90 & 6 & 20 & 91 & 53 & 58.24 &  53.31 \\ \hline
100 & 6 & 20 & 89 & 51 & 57.30 &  44.51 \\ \hline
110 & 6 & 20 & 94 & 55 & 58.51 &  81.59 \\ \hline
120 & 6 & 20 & 82 & 50 & 60.98 &  66.91 \\ \hline
130 & 6 & 20 & 85 & 54 & 63.53 &  72.28 \\ \hline
140 & 6 & 20 & 79 & 46 & 58.23 &  30.09 \\ \hline
150 & 6 & 20 & 82 & 50 & 60.98 &  54.64 \\ \hline
160 & 6 & 20 & 92 & 52 & 56.52 &  11.74 \\ \hline
170 & 6 & 20 & 87 & 54 & 62.07 &  61.94 \\ \hline
180 & 6 & 20 & 97 & 54 & 55.67 &  25.49 \\ \hline
190 & 6 & 20 & 91 & 53 & 58.24 &  19.97 \\ \hline
200 & 6 & 20 & 90 & 50 & 55.56 &  64.05 \\ \hline
210 & 6 & 20 & 96 & 55 & 57.29 &  30.46 \\ \hline
220 & 6 & 20 & 105 & 65 & 61.90 &  86.30 \\ \hline
230 & 6 & 20 & 89 & 56 & 62.92 &  85.85 \\ \hline
240 & 6 & 20 & 86 & 51 & 59.30 &  58.13 \\ \hline
250 & 6 & 20 & 101 & 62 & 61.39 &  91.45 \\ \hline
260 & 6 & 20 & 89 & 50 & 56.18 &  18.62 \\ \hline
270 & 6 & 20 & 100 & 66 & \cellcolor{black!30}\textbf{66.00} &  89.81 \\ \hline
280 & 6 & 20 & 104 & 60 & 57.69 &  63.16 \\ \hline
290 & 6 & 20 & 95 & 59 & 62.11 &  63.31 \\ \hline
300 & 6 & 20 & 105 & 63 & 60.00 &  65.96 \\ \hline
310 & 6 & 20 & 100 & 64 & 64.00 &  93.60 \\ \hline
320 & 6 & 20 & 116 & 73 & 62.93 &  71.80 \\ \hline
330 & 6 & 20 & 101 & 62 & 61.39 &  96.31 \\ \hline
340 & 6 & 20 & 117 & 71 & 60.68 &  83.96 \\ \hline
350 & 6 & 20 & 119 & 70 & 58.82 &  63.92 \\ \hline
360 & 6 & 20 & 104 & 63 & 60.58 &  36.01 \\ \hline
370 & 6 & 20 & 111 & 69 & 62.16 &  \cellcolor{black!30}\textbf{132.79} \\ \hline
380 & 6 & 20 & 110 & 65 & 59.09 &  67.83 \\ \hline
390 & 6 & 20 & 120 & 69 & 57.50 &  42.80 \\ \hline
400 & 6 & 20 & 115 & 73 & 63.48 &  85.81 \\ \hline
\hline
\end{tabular}
\end{table}

The best result for accuracy was achieved using a 270-day training window for the EMD-ELM, in which 100 negotiations are run, whereas $66\%$ with profit. The best return is generated with a window of 370 days, achieving $132.79\%$. The results using TOPSIS with the stochastic D and CCI criteria for selection are shown in Table \ref{tab-result-topsis-emd-elm2}. In this case, the best return and accuracy results are achieved using a 50-day training window, with 79 negotiations of which $65.82\%$ are profitable, generating a return of $135.89\%$. 

\begin{table}[ht]
\tiny
\centering
\caption{Results using TOPSIS (D-CCI) + EMD-ELM for the year 2016}
\label{tab-result-topsis-emd-elm2}
\begin{tabular}{|l|l|l|l|l|l|l|}
\hline
Window & Delay number & Neurons & Negotiations & Neg. with profit & Accuracy & Percentage return\\\hline
20 & 6 & 20 & 76 & 46 & 60.53 &  70.03 \\ \hline
30 & 6 & 20 & 76 & 39 & 51.32 &  -4.03 \\ \hline
40 & 6 & 20 & 78 & 42 & 53.85 &  57.32 \\ \hline
50 & 6 & 20 & 79 & 52 & \cellcolor{black!30}\textbf{65.82} &  \cellcolor{black!30}\textbf{135.89} \\ \hline
60 & 6 & 20 & 78 & 50 & 64.10 &  73.71 \\ \hline
70 & 6 & 20 & 81 & 49 & 60.49 &  76.14 \\ \hline
80 & 6 & 20 & 93 & 58 & 62.37 &  77.87 \\ \hline
90 & 6 & 20 & 93 & 55 & 59.14 &  79.28 \\ \hline
100 & 6 & 20 & 91 & 55 & 60.44 &  94.33 \\ \hline
110 & 6 & 20 & 92 & 57 & 61.96 &  116.93 \\ \hline
120 & 6 & 20 & 84 & 49 & 58.33 &  55.59 \\ \hline
130 & 6 & 20 & 87 & 56 & 64.37 &  98.80 \\ \hline
140 & 6 & 20 & 80 & 47 & 58.75 &  44.41 \\ \hline
150 & 6 & 20 & 81 & 47 & 58.02 &  64.21 \\ \hline
160 & 6 & 20 & 93 & 51 & 54.84 &  25.92 \\ \hline
170 & 6 & 20 & 80 & 48 & 60.00 &  30.51 \\ \hline
180 & 6 & 20 & 95 & 50 & 52.63 &  18.23 \\ \hline
190 & 6 & 20 & 90 & 48 & 53.33 &  0.71 \\ \hline
200 & 6 & 20 & 88 & 48 & 54.55 &  53.17 \\ \hline
210 & 6 & 20 & 90 & 51 & 56.67 &  32.20 \\ \hline
220 & 6 & 20 & 103 & 61 & 59.22 &  93.69 \\ \hline
230 & 6 & 20 & 86 & 55 & 63.95 &  84.27 \\ \hline
240 & 6 & 20 & 86 & 51 & 59.30 &  57.76 \\ \hline
250 & 6 & 20 & 100 & 59 & 59.00 &  77.59 \\ \hline
260 & 6 & 20 & 87 & 48 & 55.17 &  25.11 \\ \hline
270 & 6 & 20 & 92 & 58 & 63.04 &  86.98 \\ \hline
280 & 6 & 20 & 96 & 54 & 56.25 &  64.78 \\ \hline
290 & 6 & 20 & 92 & 53 & 57.61 &  55.10 \\ \hline
300 & 6 & 20 & 106 & 61 & 57.55 &  72.64 \\ \hline
310 & 6 & 20 & 98 & 58 & 59.18 &  82.67 \\ \hline
320 & 6 & 20 & 110 & 65 & 59.09 &  61.47 \\ \hline
330 & 6 & 20 & 98 & 58 & 59.18 &  89.72 \\ \hline
340 & 6 & 20 & 116 & 66 & 56.90 &  72.52 \\ \hline
350 & 6 & 20 & 111 & 63 & 56.76 &  74.20 \\ \hline
360 & 6 & 20 & 106 & 64 & 60.38 &  67.72 \\ \hline
370 & 6 & 20 & 108 & 63 & 58.33 &  120.21 \\ \hline
380 & 6 & 20 & 111 & 62 & 55.86 &  68.61 \\ \hline
390 & 6 & 20 & 128 & 71 & 55.47 &  54.01 \\ \hline
400 & 6 & 20 & 114 & 68 & 59.65 &  93.67 \\ \hline
\end{tabular}
\end{table}

The results using TOPSIS with all the indicators are shown in Table \ref{tab-result-topsis-emd-elm3}. TOPSIS used with four indicators as criteria presents results lower than the previous configurations. The best accuracy is achieved with a 50-day window and the best return with a 290-day window.

\begin{table}[ht]
\tiny
\centering
\caption{Results of TOPSIS (RSI-K-D-CCI) + EMD-ELM for year 2016}
\label{tab-result-topsis-emd-elm3}
\begin{tabular}{|l|l|l|l|l|l|l|}
\hline
Window & Delay number & Neurons & Negotiations & Neg. with profit & Accuracy & Percentage return\\\hline
20 & 6 & 20 & 94 & 49 & 52.13 &  7.24 \\ \hline
30 & 6 & 20 & 89 & 48 & 53.93 &  59.15 \\ \hline
40 & 6 & 20 & 87 & 45 & 51.72 &  33.60 \\ \hline
50 & 6 & 20 & 73 & 44 & \cellcolor{black!30}\textbf{60.27} &  73.43 \\ \hline
60 & 6 & 20 & 70 & 40 & 57.14 &  41.31 \\ \hline
70 & 6 & 20 & 72 & 39 & 54.17 &  20.03 \\ \hline
80 & 6 & 20 & 79 & 45 & 56.96 &  51.59 \\ \hline
90 & 6 & 20 & 93 & 51 & 54.84 &  25.76 \\ \hline
100 & 6 & 20 & 90 & 48 & 53.33 &  26.24 \\ \hline
110 & 6 & 20 & 95 & 53 & 55.79 &  23.51 \\ \hline
120 & 6 & 20 & 82 & 43 & 52.44 &  16.83 \\ \hline
130 & 6 & 20 & 81 & 47 & 58.02 &  27.35 \\ \hline
140 & 6 & 20 & 80 & 42 & 52.50 &  9.70 \\ \hline
150 & 6 & 20 & 76 & 43 & 56.58 &  45.47 \\ \hline
160 & 6 & 20 & 96 & 48 & 50.00 &  6.92 \\ \hline
170 & 6 & 20 & 74 & 38 & 51.35 &  23.89 \\ \hline
180 & 6 & 20 & 81 & 40 & 49.38 &  1.30 \\ \hline
190 & 6 & 20 & 78 & 38 & 48.72 &  18.14 \\ \hline
200 & 6 & 20 & 77 & 41 & 53.25 &  75.34 \\ \hline
210 & 6 & 20 & 91 & 47 & 51.65 &  22.04 \\ \hline
220 & 6 & 20 & 98 & 53 & 54.08 &  59.66 \\ \hline
230 & 6 & 20 & 87 & 50 & 57.47 &  78.37 \\ \hline
240 & 6 & 20 & 77 & 43 & 55.84 &  60.29 \\ \hline
250 & 6 & 20 & 86 & 47 & 54.65 &  38.54 \\ \hline
260 & 6 & 20 & 81 & 43 & 53.09 &  14.08 \\ \hline
270 & 6 & 20 & 88 & 53 & 60.23 &  51.19 \\ \hline
280 & 6 & 20 & 89 & 48 & 53.93 &  57.71 \\ \hline
290 & 6 & 20 & 84 & 50 & 59.52 &  \cellcolor{black!30}\textbf{80.60} \\ \hline
300 & 6 & 20 & 101 & 59 & 58.42 &  69.97 \\ \hline
310 & 6 & 20 & 100 & 54 & 54.00 &  50.70 \\ \hline
320 & 6 & 20 & 100 & 54 & 54.00 &  22.17 \\ \hline
330 & 6 & 20 & 101 & 57 & 56.44 &  56.97 \\ \hline
340 & 6 & 20 & 111 & 57 & 51.35 &  -2.27 \\ \hline
350 & 6 & 20 & 112 & 56 & 50.00 &  11.68 \\ \hline
360 & 6 & 20 & 104 & 55 & 52.88 &  10.02 \\ \hline
370 & 6 & 20 & 106 & 58 & 54.72 &  66.07 \\ \hline
380 & 6 & 20 & 110 & 60 & 54.55 &  23.19 \\ \hline
390 & 6 & 20 & 119 & 63 & 52.94 &  7.05 \\ \hline
400 & 6 & 20 & 117 & 66 & 56.41 &  23.90 \\ \hline

\end{tabular}
\end{table}

\subsection{Results for the year 2017}

TOPSIS was applied based on the best criteria found for the year 2016. The configurations used were the ones with the best return and the best accuracy. In addition, TOPSIS was also used with all indicators, the random strategy and the Bovespa index for comparison. The results are shown in Table \ref{tab-topsis-2017}.

\begin{table}[ht]
\centering
\caption{Results obtained for the year 2017}
\label{tab-topsis-2017}
\begin{tabular}{|l|l|l|l|}
\hline
Method    & Criteria & Accuracy & Percentage return \\\hline
TOPSIS& Stoch. D, CCI                    &55.10\% & 44,17\%\\\hline
TOPSIS& Stoch. K, est. D, CCI            &\textbf{55.92\%} & \textbf{86.31\%}\\\hline
TOPSIS& all &51.84\% & 43,97\%\\\hline
Random & - & $51.06\pm2.75\%$ & $30.50\pm35.53\%$\\\hline  
Ibovespa  & - & -              & 28.21\%\\ \hline
\end{tabular}
\end{table}

\begin{figure}[ht]
    \centering
    \includegraphics[width=1\textwidth]{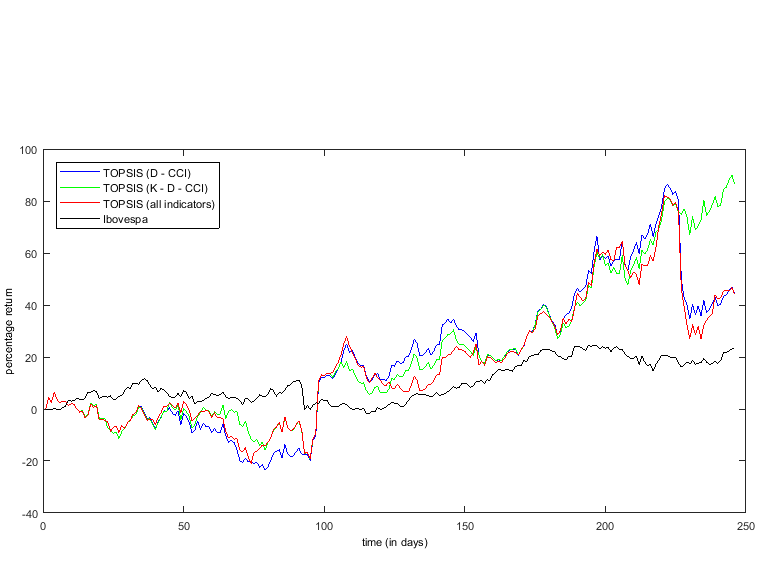}
    \caption{Percentage return obtained for the year 2017}
    \label{fig:retorno}
\end{figure}

As in the previous year, the selection made by TOPSIS outperformed the random selection and the Bovespa index. The configuration with indicators stochastic D, CCI and stochastic K presented the best result. TOPSIS using stochastic D and CCI that presents the best return for the year 2016 was not able to maintain the best performance for 2017. The performance of the methods, except the random one, can be observed throughout the year in terms of return as shown in Figure \ref{fig:retorno}.

Next, to the TOPSIS is added the EMD-ELM. The first configuration uses TOPSIS with indicators stochastic K and D and CCI as criteria and the EMD-ELM using a training window with 270 samples, 6 delays and 20 neurons. The second configuration uses a TOPSIS with stochastic indicators stochastic D and CCI as criteria and EMD-ELM with a training window of 50, 6 delays and 20 neurons. The results are shown in Table \ref{tab-topsis-emd-elm-2017}.

\begin{table}[ht]
\small
\centering
\caption{Results with TOPSIS+EMD-ELM for the year 2017 }
\label{tab-topsis-emd-elm-2017}
\begin{tabular}{|l|l|l|l|l|l|l|}
\hline
Criteria & window- & Negotiations & Negotiations &Accuracy & Percentage \\
 & delay- &  & with & & return \\
  & neurons &  & profit & &  \\\hline
K-D-CCI & 270-6-20 & 96 & 57 & 59.38\% &  49.28\% \\ \hline
D-CCI & 50-6-20    & 80 & 44 & 55\% &  4.87\% \\ \hline
\end{tabular}
\end{table}

The first configuration presented the best results both in terms of return and accuracy. By adding EMD-ELM, there is an increase in the percentage of negotiations with profit, however the return decreases.

\subsection{Results analysis}

In most cases for the year 2016, TOPSIS was able to overcome both the random selection and the Bovespa index, both in accuracy (except in the Bovespa index, for which it is not possible to calculate accuracy) and in return. The configuration with the best accuracy were: stochastic D, stochastic K and CCI, reaching $58.30\%$. For return, the best configuration uses stochastic D and CCI, achieving $326.77\%$. In the various settings of criteria, only in one that does not outperform the Bovespa index. It is important to note that, not always the best accuracy yields the best return. A low-profit selection is counted as a hit in the same way as another with a high profit. The choice of the correct criteria has a great influence on the result of the selection made by TOPSIS. By analysing the obtained results it is possible to note that some indicators seem to generate better results when used as is the case of CCI. Based on the evaluation of the various combinations of indicators for the year 2016, it was possible to choose good configurations for the year 2017. TOPSIS continues presenting better results in terms of accuracy and return. The return in 2017 however was not so  high as in 2016, but it is still considerable.

Next, to TOPSIS is added EMD-ELM, which acts by confirming whether the stock selected by TOPSIS should be purchased. Therefore, the EMD-ELM sometimes refuses to buy a stock and, consequently, this reduces the number of negotiations. However, this rejection is not always correct, thus reducing the number of profit-making operations. So, the return was lower, but the reliability of operations increased. A key factor for the EMD-ELM was to determine the size of the training window. This parameter is very important because the trend detected by the EMD for a month may be different from that of a year for example. Several settings for the size of the training window have been investigated for the TOPSIS with the three configurations previously tested in the year 2016.

The highest accuracy was achieved using the TOPSIS with stochastic K and stochastic D and CCI and a training window of 270 days. In this case EMD-ELM was able to increase the accuracy of $58.30\%$ (using only TOPSIS) to $66\%$ . The return however droped from $269.73\%$  to $89.81\%$. The best return was obtained through the use of TOPSIS with stochastic D and CCI criteria and a 50 day window. However, in relation to the use only of TOPSIS, there was a decrease from $326.77\%$  to  $132.79\%$. However, the accuracy increased from $57.89\%$ to $65.82\%$. Using the two settings cited above for the year 2017, the behaviour remained similar to 2016 in relation to using only TOPSIS. Using the EMD-ELM with the 270-day window, the accuracy rises from $55.92\%$ to  $59.3\%$ and profit falls from $86.31\%$  to $49.28\%$. Already using the second configuration, with a window of 50 days, the accuracy remains practically the same dropping from  $55.1\%$ to  $55\%$ and the return falls from $44.17\%$ to  $4.87\%$. 

A direct comparison with results of similar work is very difficult, if not impossible, due to the fact that they use different scenarios and approaches. However, in order to refer the obtained results in this work, we mention some of them. Vargas, Lima and Evsukoff \cite{vargas2017} used deep learning methods to predict the direction of movement of the S\&P500 index using as input technical indicators and news headlines. The methodology achieved  $62.03\%$ of hits. Nelson, Pereira and Oliveira \cite{nelson2017} used Long-Short term memory (LSTM) to predict the trend of stock prices in the Brazilian market based on historical  prices and technical indicators. They used the time interval of 15 minutes, and achieved an average accuracy of $55.9\%$. Kara, Boyacioglu and Baykan \cite{kara2011} used artificial neural networks and support vector machine (SVM) to predict the ISE (Istanbul Stock Exchange) movement. In the experimental results the neural network achieved on average $75.74\%$ and the SVM  $71.52\%$ of accuracy.

\section{Conclusions}
\label{sec-conclusions}

In this work a hybrid method was developed for stock trading combining TOPSIS, EMD, and ELM. The method was built starting with the TOPSIS technique only, and then adding the EMD-ELM hybrid model. TOPSIS acts by selecting each day the most suitable stock to buy among a set of stocks according the technical indicators. The EMD-ELM hybrid model acts by confirming or refusing the purchasing order. The EMD decomposes the series, extracting the trend component, and the ELM classifies the trend whether it is high or low.  The method was tested in the Brazilian market for an environment of 50 stocks. The data were collected for the years 2016 and 2017. The 2016 data was used for training the method parameters. The metrics used were the percentage of trades with profit and the return generated. The selection made by TOPSIS was compared to the random selection and the Bovespa index.  TOPSIS achieved good results for 2016, outperforming the other two methods. Nonetheless, the technique is strongly influenced by the criteria (technical indicators) used. Based on the indicators used as criteria with data of the year 2016, the technique continued to present good results for 2017. The confirmation of the purchase by EMD-ELM increased the percentage of negotiations with profit, but reduced the total number of negotiations. The method also presented a reduction in return. The size of the training sliding window has great influence on the method performance. The proposed model combining  TOPSIS with EMD-ELM is able to make decision of buying a stock at each end of the day turns out to be quite promising in terms of profit and return. So, future work may incorporate other kind of neural networks or even fuzzy logic.  Another interesting issue is when the EMD-ELM is used to confirm the purchase of stocks, fewer trades are made, increasing the time between one purchase and another. In this work, sales of the stock selling occur at the end of the day after the purchase. However, a decision-making system to sell would be interesting in this situation because it might set a more suitable time to sell.

\section*{Acknowledgement}
E. Ebermam thanks the scholarship of the Brazilian agency
CNPq. R.A. Krohling would like to thank the Brazilian agency
CNPq and the local Agency of the state of Espirito Santo
FAPES for financial support under grant No. 309161/2015-0
and No. 039/2016, respectively.

\section*{References}

\bibliography{mybibfile}

\appendix
\section{List of stocks considered in this work}
\label{appendix-a}

\begin{table}[ht]
\centering
\caption{Codes and names of the stocks considered in this work}
\label{tab-acoes}
\begin{tabular}{|l|l|l|l|}
\hline
Code    & Stocks & Code  & Stocks\\\hline
 ABEV3 & Ambev SA      & GOAU4 & Gerdau Met \\\hline
 BBAS3 & Brasil        & HYPE3 & Hypera\\\hline
 BBDC3 & Bradesco ON   & ITSA4 & Itaúsa\\\hline
 BBDC4 & Bradesco PN   & ITUB4 & Itaú Unibanco\\\hline
 BBSE3 & BB Seguridade & JBSS3 & JBS\\\hline   
 BRAP4 & Bradespar     & KROT3 & Kroton\\\hline
 BRFS3 & BRF SA        & LAME4 & Lojas Americanas\\\hline
 BRKM5 & Braskem       & LREN3 & Lojas Renner\\\hline
 BRML3 & BR Malls Par  & MRFG3 & Marfrig\\\hline
 BVMF3 & B3            & MRVE3 & MRV\\\hline
 CCRO3 & CCR SA        & MULT3 & Multiplan\\\hline
 CIEL3 & Cielo         & NATU3 & Natura\\\hline
 CMIG4 & Cemig         & PCAR4 & P. Açúcar CDB\\\hline
 CPFE3 & CPFE Energia  & PETR3 & Petrobras ON\\\hline
 CPLE6 & Copel         & PETR4 & Petrobras PN\\\hline
 CSAN3 & Cosan         & QUAL3 & Qualicorp\\\hline
 CSNA3 & Sid. Nacional & RADL3 & Raia Drogasil\\\hline
 CYRE3 & Cyrela Realt  & SANB11& Santander BR\\\hline
 ECOR3 & Ecorodovias   & SBSP3 & Sabesp\\\hline
 EMBR3 & Embraer       & TIMP3 & Tim Part SA\\\hline
 ENBR3 & Energias BR   & UGPA3 & Ultrapar\\\hline
 EQTL3 & Equatorial    & USIM5 & Usiminas\\\hline
 ESTC3 & Estácio Part  & VALE3 & Vale\\\hline
 FIBR3 & Fibria        & VIVT4 & Telef Brasil\\\hline
 GGBR4 & Gerdau        & WEGE3 & Weg\\\hline
\end{tabular}
\end{table}

\end{document}